\documentstyle[12pt]{article} 
\begin{document} 
\thispagestyle{empty} 
\begin{flushright}
UA/NPPS-7-1998
\end{flushright}
\begin{center}
{\large{\bf PROBING THE BOUNDARIES OF THE HADRONIC PHASE\\ 
THROUGH A STRANGENESS INCLUDING \\ 
STATISTICAL BOOTSTRAP MODEL (S-SBM) \\}} 
\vspace{2cm} 
{\large A. S. Kapoyannis, C. N. Ktorides and A. D. Panagiotou}\\ 
\smallskip 
{\it University of Athens, Division of Nuclear and Particle Physics,\\ 
GR-15771 Athina, Hellas}\\ 
\vspace{1.0cm} 
{\it Submitted in Physical Review D}
\vspace{1.0cm}
\end{center}
\begin{abstract} 
A recently constructed strangeness-including Statistical Bootstrap Model 
(S-SBM), which defines the limits of the hadronic phase and provides for a 
phase beyond, is further extended so as to include a factor $\gamma_s$ that 
describes strangeness suppression. The model is then used to analyse the 
multiplicity data from collision experiments in which the colliding entities 
form isospin symmetric systems, the primary focus being on $S+S$ 
interactions (NA35 collaboration). An optimal set of thermodynamical 
variables is extracted through a fit to both the inclusive $4\pi$ and 
midrapidity data. 
The assumption that the measured particles originate from a thermally and 
partial-chemically equilibrated source described by the S-SBM is 
satisfactorily established. 
The proximity of the thermodynamical variables extracted from the $S+S$ 
data to the limits of the hadronic phase is systematically investigated. 
Finally, experimental data from $p+\overline{p}$ collisions (UA5 
collaboration) are similarly analysed. 
\end{abstract} 
\vspace{1cm}
\newpage
\setcounter{page}{1} 
 
{\large{\bf 1. Introduction}} 
 
The description of multiparticle hadronic states which emerge from 
a collision process at very high energies can, most 
economically, be conducted by adopting thermalization hypotheses 
pervading their evolution course. It is, in 
fact, the quintessence of the statistical mechanical approach to a given 
system with a large number of degrees of freedom that it transcribes a 
hopelessly complicated microscopic description into a corresponding 
concise one which employs a handful of thermodynamical variables (e.g. 
temperature, volume, chemical potentials and the like). 
 
Naturally, the ultimate goal of any attempt to address issues involving 
multiparticle production in high energy collisions is to account for the 
dynamics which give rise to the experimentally recorded profile of the final 
state of the system. For the particular case of heavy ion collisions, on which 
the bulk of our attention will be focused in this work, the 
question of fundamental interest is whether the overall collision process has 
gone through a deconfined phase and, if yes, in what way 
can one pick signatures of this occurrence in the composition of the final 
state. 
 
Given the above general remarks, let us now become more concrete and 
consider the basic thermodynamical scenario which has 
been adopted by the pioneers of this approach [1-3]. 
To describe a system that has been produced by one (or more) high energy 
density source(s), one assumes a dynamical evolution that 
reaches freeze-out points (thermal, chemical) at which corresponding 
thermodynamical quantities acquire their final equilibrium values, 
reflected in the observed particle species and multiplicities. 
 
The statistical mechanical analysis evaluating the experimental data can 
be conducted in the spirit: (a) Taking for granted the validity of the 
thermal picture and looking for possible discrepancies which provide 
signals of something interesting having taken place, (b) testing the 
reliability of a given thermal model, (c) a combination thereof. In any case, 
it is particle multiplicities (or ratios) which provide concrete numerical 
input for the theoretical analysis. Such information addresses itself to the 
chemical composition of the system, which means that chemical potentials 
become a key ingredient in this type of description. 
In this context, one must also consider the 
prospect of amending the equilibrium scheme by inserting 
parameters which account for {\it partial} equilibrium conditions with 
respect to a given quantity. This possibility will, in fact, 
prominently enter the present analysis in connection with saturation of 
strangeness phase space. 
 
Theoretical confrontations of thermal hadron production in various 
types of experiments, ranging from $e^+ +e^-$ to nucleus-nucleus 
collisions, have been conducted within the framework of a statistical model 
corresponding to an ideal hadron gas with results that, {\it a posteriori} at 
least, justify the premises of the whole approach [1-11]. However, such a 
model has a built-in assumption that the hadronic world has no bounds and it 
does not anticipate the existence of a new phase, beyond the hadronic one, 
when a critical point in the space of thermodynamical parameters is reached. 
By contrast, the statistical bootstrap model (SBM), originally introduced by 
Hagedorn [12], see also [13-16], {\it does} anticipate the end of the 
hadronic domain and, under certain conditions on model parameters [17], is 
consistent with the existence of a new phase beyond the hadronic one. Given 
this state of affairs, as well as the fact that the quantum number of 
strangeness plays a central role in revealing the quark-gluon plasma 
(QGP) phase, we have recently constructed an extension of the SBM so as 
to incorporate Strangeness [18,19], 
henceforth to be referred to as S-SBM. The model exhibits attractive 
properties consistent with what one anticipates from the transition 
between the hadronic and the new, presumably QGP, phase of matter. 
 
At present, the S-SBM treats $u$ and $d$ quarks 
on the same footing so that a single fugacity corresponds to both of them. 
This amounts to considering only isospin symmetric systems, i.e. systems 
for which the number of $u$-quarks minus $\bar{u}$-quarks is held equal 
to the number of $d$-quarks minus $\bar{d}$-quarks. Then it is easy to 
prove that 
\begin{equation} 
n_u-n_{\bar{u}}=n_d-n_{\bar{d}}\Leftrightarrow 
\lambda_u=\lambda_d\equiv\lambda_q\;\;, 
\end{equation} 
where the $n$'s stand for quark numbers and the $\lambda$'s for 
corresponding fugacities\footnote{Our notational conventions regarding 
fugacity labels will be specified in the next section.}. 
The above relation, in the presence of the condition that the total 
strangeness vanishes, is equivalent to 
\begin{equation} 
B=2Q\Leftrightarrow\lambda_Q=1\;\;, 
\end{equation} 
where $Q$ is the total electric charge and $\lambda_Q$ is the 
corresponding fugacity. So in the aforementioned situation we need not 
consider the existence of the additional fugacity $\lambda_Q$ and the present 
form of S-SBM ascribes precisely to these systems. 
 
In this paper we intend to confront, primarily, experimental data coming 
from isospin symmetric $S+S$ collisions (NA35 experiment at CERN) 
within the framework of the S-SBM. For purposes 
of completeness, we shall also use our methodology for the 
evaluation of $p+\overline{p}$ data (UA5 experiment at CERN). 
We shall be addressing ourselves to both hadron 
multiplicities and corresponding particle ratios which involve strange 
particles, assuming partial strangeness saturation. To this 
end, we shall introduce into our scheme the additional parameter $\gamma_s$, 
which will be disscused in the following section and can be 
identified as the ``fugacity'' pertaining to the number of strange plus 
antistrange quarks. 
 
Our analysis will cover both the inclusive 4$\pi$ set of data and the, 
more restricted, midrapidity region. A fine tuning due to 
corrections from Bose/Fermi statistics will also be taken into account, given 
that the SBM scheme adopts the Boltzmann distribution (maximum entropy 
content). Our purpose is twofold: first to find out whether a thermal 
description based on the S-SBM accounts for the experimental data in a 
satisfactory manner, and second to find the proximity of the 
data to the critical surface which sets the limits of the hadronic phase. 
 
The organisation of the paper is the following. In the next section we shall 
consider the ramifications brought by the inclusion of the $\gamma_s$ 
parameter into the S-SBM scheme. In Section 3 we shall proceed to 
present the methodology by which we propose to conduct our analysis of 
observed particle multiplicities. The actual examination of the 
experimental data, both for the $S+S$ and 
$p+\overline{p}$ cases, will be presented in Section 4. A $\chi^2$ fit will be 
employed to asses our theoretical predictions. We shall devote the last 
section to a discussion-evaluation of our results, 
paying special attention to the issue of whether and/or how close the $S+S$ 
data come to revealing the attainment of the QGP phase, always according 
to our model. Two appendices (A and B) are devoted to technical 
matters concerning routines used in our numerical calculations, while 
in appendix C we discuss the particle ratios vs particle multiplicities issue.

\newpage 
{\large{\bf 2. Partial strangeness equilibrium in the S-SBM}} 
 
We shall begin this section by presenting, in an outline form, the S-SBM 
construction of Ref. [18,19]. The underlying assumption in that work was that 
the dynamical development of the hadronic system formed after a high 
energy collision is characterised by chemical equilibrium throughout. 
Subsequently, we shall consider the generalisation of this scheme brought 
about by allowing for only {\it partial} strangeness equilibrium. 
 
The two basic ingredients going into any SBM scheme are the following: 
 
(a) The so-called {\it bootstrap equation}, which incorporates 
in a self-consistent manner the internal dynamics operating in a relativistic 
multiparticle system. 
 
(b) A statistical mechanical account of the said system in terms 
of a suitably defined partition function (equilibrium mode of description) 
 
The bootstrap equation refers to a {\it spectrum function} and receives 
input from the set of all known hadrons through the so-called {\it input 
function}. Relativistic invariance in the casting of the system leads to a 
term of the form $B(p^2)\tau(p^2,b,s)$, where $B(p^2)$ is a kinetic 
factor and $\tau(p^2,b,s)$ the naturally defined spectrum function which 
exhibits dependence on baryon $(b)$ and strangeness $(s)$ numbers. 
Rearrangements of the kind 
\begin{equation} 
B(p^2)\tau(p^2,b,s)={\tilde B}(p^2){\tilde\tau}(p^2,b,s) 
\end{equation} 
define different versions of the bootstrap scheme depending on whether the 
${\tilde \tau}$-function carries all or part of the dynamics acting 
internally. 
In the latter case, the $\tilde B$-factor can also assume a dynamical role. 
 
The particular choice specifying the S-SBM has been extensively discussed 
in [18,19], on the basis of the decisive physical advantages it exhibits 
[20], is 
\begin{equation} 
B(p^2)=\frac{2V^\mu p_\mu}{{(2\pi)}^3}=\frac{2Vm}{{(2\pi)}^3}\;, 
\end{equation} 
where the last expression refers to the rest frame of the 
particle/fireball. The above selection simply implies that $B(p^2)$, which 
takes the form $B(m^2)$ in the rest frame, remains a purely kinematical 
quantity. 
 
Setting $B(m^2)=H_0m^2$ one arrives at the notable relation 
\begin{equation} 
H_0=\frac{2}{{(2\pi)}^3 4B} \label{bb} 
\end{equation} 
which links the bootstrap scheme parameter $H_0$ with the MIT bag 
constant $B$, the latter entering through the relation 
$\frac{\textstyle 1}{\textstyle 4B}=\frac{\textstyle V}{\textstyle m}$. 
It turns out that $H_0$ is directly proportional to the maximum value of 
the critical temperature $T_0$. 
 
A Laplace transformation which takes the set of variables $(p_\mu,b,s)$ to 
the set $(\beta_\mu,\lambda_B,\lambda_S)$ introduces the inverse 
four-temperature, along with baryon and strangeness fugacities, into the 
bootstrap scheme. One is thereby led to the construction of a partition 
function \linebreak $Z(\beta,V, \lambda_B,\lambda_S)$ in the fireball rest 
frame. In this way a statistical mechanical description of the system is 
introduced, giving the final form to the S-SBM. 
 
The analysis conducted in Ref. [18,19], with respect to the S-SBM, took 
place in the 3-dimensional space of the thermodynamical variables 
$(T,\mu_q,\mu_s)$, where $\mu_q$ and $\mu_s$ stand for up-down and 
strange quark chemical potentials. In this space, the following two surfaces 
were considered: 
\newline (a) The {\it critical surface} specified by one of the conditions 
\begin{equation} 
\varphi(T_{cr},\mu_{q\;cr},\mu_{s\;cr};H_0)=\ln 4- 
1,\;\;\;\;G(T_{cr},\mu_{q\;cr},\mu_{s\;cr};H_0)=\ln 2\;, 
\end{equation} 
where $\varphi$ corresponds to the input and $G$ to the mass 
spectrum-containing function entering the bootstrap equation. This surface 
constitutes the earmark of the bootstrap 
scheme, which separates it from the ideal hadron gas model. It signifies the 
termination of the hadron phase, since the bootstrap equation does not 
posses a physically meaningful solution beyond the critical surface. In other 
words, the critical surface {\it sets the boundaries} of the hadronic world in 
the space $(T,\mu_q,\mu_s)$. 
\newline (b) The {\it $<S>=0$ surface} which imposes a strangeness 
neutrality assumption on the S-SBM construction and is specified by 
\begin{equation} 
\int_{\beta}^{\infty} x^3 \frac{\partial G(x,\lambda_q,\lambda_s;H_0)} 
{\partial \lambda_s}dx\;=\;0\;\;. 
\end{equation} 
 
As already mentioned in the introduction, our aim in this paper is to 
extend the framework of the S-SBM so as to allow for partial strangeness 
equilibrium before confronting experimental data (particle multiplicities or 
ratios). This is most conveniently done by introducing the variable 
$\gamma_s$ which is, in fact, a fugacity related to the number of 
$s$-quarks plus $\bar{s}$-quarks [21] (we shall henceforth call this 
number $|s|\equiv n_s+n_{\bar{s}}$): 
\begin{equation} 
\gamma_s\equiv\lambda_{|s|}=\exp (\mu_{|s|}/T)\;, 
\end{equation} 
 
The bootstrap equation in this generalised model reads ($g_{bs|s|}$ denotes 
degeneracy factors applicable to the given set of labels) 
\[\tilde{B}(p^2)\tilde{\tau}(p^2,b,s,|s|)= 
\underbrace{g_{bs|s|}\tilde{B}(p^2)\delta_0(p^2- 
m_{bs|s|}^2)}_{input\;term} 
+\sum_{n=2}^\infty\frac{1}{n!}\int\delta^4(p-\sum_{i=1}^np_i)\cdot\] 
\begin{equation} 
\cdot\sum_{\{b_i\}}\delta_K(b-\sum_{i=1}^nb_i) 
\sum_{\{s_i\}}\delta_K(s-\sum_{i=1}^ns_i) 
\sum_{\{|s|_i\}}\delta_K(|s|-\sum_{i=1}^n|s|_i) 
\prod_{i=1}^n\tilde{B}(p_i^2)\tilde{\tau}(p_i^2,b_i,s_i,|s|_i)d^4p_i\;. 
\end{equation} 
In the above equation, baryon number is denoted by ``$b$'' and 
strangeness by ``$s$''. Then we can perform in (9) four Laplace 
transformations which lead to the replacement (after going to the rest 
frame of the system) 
\begin{equation} 
(p^\mu,b,s,|s|)\longrightarrow(\beta,\lambda_B,\lambda_S,\gamma_s)\, 
\end{equation} 
where $\lambda_B$ and $\lambda_S$ are the fugacities corresponding to 
baryon number and strangeness, respectively. 
Since we are accustomed to working with fugacities corresponding to 
quarks, we can equivalently use the set defined by 
\begin{equation} 
\lambda_q=\lambda_B^{1/3}\;,\;\;\;\; 
\lambda_s^{neq}=\gamma_s\lambda_B^{1/3}\lambda_S^{-1}\;,\;\;\;\; 
\lambda_{\overline{s}}^{neq}=\gamma_s\lambda_B^{-1/3}\lambda_S\;. 
\end{equation} 
The up and down quark fugacity is denoted by $\lambda_q$, whereas 
$\lambda_s^{neq}$ and $\lambda_{\overline{s}}^{neq}$ denote the $s$- 
quark and $\overline{s}$-quark fugacites respectively. The index $neq$ 
means that the chemical equilibrium connected with strangeness is, 
in principle, not acheived. 
The factor $\gamma_s$ entering (11) can, then, be given by 
\begin{equation} 
\gamma_s^2=\lambda_s^{neq}\lambda_{\overline{s}}^{neq}\;. 
\end{equation} 
In order to stay in agrement with the conventions of the quark fugacities 
used in [18,19], as well with the ones used by other authors, we shall 
employ the set $(\lambda_q,\lambda_s,\gamma_s)$. Now 
$\lambda_s\equiv\lambda_B^{1/3}\lambda_S^{-1}$ is the fugacity of the 
$s$-quark which corresponds to the chemical equilibrium of strangeness. 
 
The bootstrap equation acquires the form 
\begin{equation} 
\varphi(T,\lambda_q,\lambda_s,\gamma_s)= 
2G(T,\lambda_q,\lambda_s,\gamma_s)-\exp[ 
G(T,\lambda_q,\lambda_s,\gamma_s)]+1\;, 
\end{equation} 
with the functions $\varphi$ and $G$ given by 
\begin{equation} 
\varphi(T,\lambda_q, \lambda_s, \gamma_s;H_0)=2\pi H_0T 
\sum_{\rm a}\lambda_{\rm a} 
\sum_i g_{{\rm a}i}m_{{\rm a}i}^3K_1 
(m_{{\rm a}i}/T)\;, 
\end{equation} 
\begin{equation} 
G(T,\lambda_q,\lambda_s, \gamma_s;H_0)=2\pi H_0 T\int_0^{\infty} 
m^3\tau_0(m^2, \lambda_q, \lambda_s, \gamma_s)K_1(m/T)dm^2\;, 
\end{equation} 
where the subscript ``$_0$'' on $\tau$ signifies the S-SBM choice given by 
eq. (4). The fugacity $\lambda_{\rm a}$ corresponds to the existing hadronic 
families that are used as input in (14) (light unflavoured mesons, kaons, 
$N\;\&\;\Delta$, $\Lambda\;\&\;\Sigma$, $\Xi$ and $\Omega$ Baryons). It 
runs over all particles and antiparticles and obeys the equality 
\begin{equation} 
\lambda_{\rm a}=\gamma_s^{n_s+n_{\bar{s}}} 
\lambda_q^{n_q-n_{\bar{q}}}\lambda_s^{n_s-n_{\bar{s}}}\;, 
\end{equation} 
where $n_i$ corresponds to the number of the $i$th flavour quark which is 
included in the ``$\rm a$'' hadron. For the case of the light unflavoured 
mesons which have the quark content of the form $c_1 q\bar{q}+c_2 s\bar{s}$ 
($c_1+c_2=1$) the corresponding fugacity equals 
\begin{equation} 
\lambda_{\rm a}= c_1 +c_2 {\gamma_s}^2\;. 
\end{equation} 
For the evaluation of the coefficients $c_1$ and $c_2$ we have used [22]. 
At the same time the partition function is amended by the addition of the 
fugacity variable $\lambda_{|s|}$ or, equivalently, the chemical potential 
$\mu_{|s|}$. 
 
The critical and $<S>=0$ surfaces are now determined by equations 
similar to those given by (6). The only difference is that $\gamma_s$ is 
also included as a variable: 
\begin{equation} 
\varphi(T_{cr},\mu_{q\;cr},\mu_{s\;cr},\gamma_{s\;cr};H_0)=\ln 4- 
1,\;\;\;\;G(T_{cr},\mu_{q\;cr},\mu_{s\;cr},\gamma_{s\;cr};H_0)=\ln 2\;, 
\end{equation} 
and 
\begin{equation} 
\int_{\beta}^{\infty} x^3 \frac{\partial 
G(x,\lambda_q,\lambda_s,\gamma_s;H_0)} 
{\partial \lambda_s}dx\;=\;0\;\;. 
\end{equation} 
 
For the S-SBM, as constructed in Refs [18,19], $T_0$, the critical temperature 
for vanishing chemical potentials, can be taken as the only free 
thermodynamical parameter of the model, 
in the place of $H_0$. $T_0$ is determined by the relation 
\begin{equation} 
\varphi(T_0,\lambda_q =1,\lambda_s =1;H_0)=\ln 4-1\;, 
\end{equation} 
and, as can be seen, is function of $H_0$ only.

In the present case the last equation becomes 
\begin{equation} 
\varphi(T_0,\lambda_q =1,\lambda_s =1,\gamma_s;H_0)=\ln 4-1\;, 
\end{equation} 
and $T_0$ is a function not only of $H_0$, but of $\gamma_s$ 
as well. It should be noted, on the other hand, that the bag constant $B$ 
continues to be in one to one correspondence with $H_0$, as can be seen 
from (5). 
 
The dependence of $T_0$ on $\gamma_s$ 
complicates the numerical analyses needed to determine the profile of the 
critical surface in the considered extension of the S-SBM. Nevertheless, the 
relevant study has been carried out and the results are presented in a 
series of figures. 
 
Figure 1a depicts two characteristic intersections of the critical surface. 
One is with the $\mu_s=0$ MeV and the other is with the $\mu_s=500$ MeV 
plane, for different values of $\gamma_s$. The unphysical value 
$\gamma_s=1.5$\footnote{The strangeness production is usually found to be 
suppressed, so physical values for $\gamma_s$ should be less than 1.} has 
been chosen to show the effect of the increase of $\gamma_s$ beyond unity. 
Figure 1b depicts the same situation as the previous one, but the 
intersections of the critical surface are with the planes $\mu_q=0$ MeV 
and $\mu_q=300$ MeV. As it can be seen, a decrease in the value of 
$\gamma_s$ below 1 leads to an expansion of the critical surface. The 
opposite happens for the increase of $\gamma_s$ above 1. 
Figure 2 shows the variation of the critical temperature $T_0$ with 
$\gamma_s$. The decrease of $\gamma_s$ leads to the increase of $T_0$ 
and vice versa. Figure 3 displays the $<S>=0$ surface for the S-SBM and 
for the ``ideal hadron gas''. The corresponding merging with the critical 
surface, in the S-SBM case, for two characteristic values of $\gamma_s$ 
(1 and 0.5) is also depicted. The above curves are plotted for constant 
fugacity $\lambda_q$ ($\mu_q=0.4$). 
 
In figures 1-3 a specific value of $H_0$ has been used which leads 
to $T_0=183$ MeV for $\gamma_s=1$, corresponding to the maximum 
value of the bag constant, $B^{1/4}=235$ MeV, according to [19]. By comparison, 
figures 4a and 4b present the intersections of the $<S>=0$ with the critical 
surface for two values of $H_0$ (one leads to $T_0=150$ MeV and the 
other to $T_0=183$ MeV, for $\gamma_s=1$). Fig. 4a shows the projections 
of the aforementioned intersections on the $\mu_s-\mu_q$ plane, while 
Fig. 4b the projections on the $\mu_s-T$ plane. 
 
In conclusion, the extension of the S-SBM to allow for 
partial strangeness equilibration {\it is} feasible and its impact on the 
model is simply to shift the critical and $<S>=0$ surfaces as $\gamma_s$ 
changes values. Its inclusion in the S-SBM scheme provides an extra 
(thermodynamical) degree of freedom, widening its scope of 
applicability.

\vspace{2cm} 
{\large{\bf 3. Methodology for Examining Experimental Data}} 
 
Our confrontation of experimental data has as its focal point particle 
multiplicities. Particle ratios will also be studied at a subsequent stage. 
Our task will be to determine the thermodynamic parameters 
$(V,T,\lambda_q,\lambda_s,\gamma_s)$ which best fit the 
($N$) particle multiplicities measured in a given experiment. 
In principle, this is equivalent to determining the parameters 
$(T,\lambda_q,\lambda_s,\gamma_s)$ which best fit a group of ($N-1$) 
indepedent particle ratios we can form from the $N$ measured 
multiplicities. Although more complicated, we shall prefer the first case, 
where it is possible, for the reasons discussed in Appendix C. 
 
Introducing hadron fugacities $\lambda_i$ into our partition function via 
an extension of the form 
\begin{equation} 
Z(V,\beta,\lambda_q,\lambda_s,\gamma_s)\;\;\rightarrow\;\; 
Z(V,\beta,\lambda_q,\lambda_s,\gamma_s,\ldots,\lambda_i,\ldots)\;\;, 
\end{equation} 
we have 
\begin{equation} 
N_i^{thermal}=\left.\left(\lambda_i\frac 
{\partial \ln 
Z(V,\beta,\lambda_q,\lambda_s,\gamma_s,\ldots,\lambda_i,\ldots)} 
{\partial \lambda_i}\right)\right|_{\ldots=\lambda_i=\ldots=1} \;\;, 
\end{equation} 
or equivalently 
\begin{eqnarray} 
\lefteqn{N_i^{thermal}(V,T, \lambda_q,\lambda_s,\gamma_s)} 
\nonumber\\ 
&&=\frac{VT^3}{4\pi^3 H_0} 
\int_0^T \frac{1}{y^5} \frac{1}{2-\exp 
[G(y,\lambda_q,\lambda_s,\gamma_s)]} \left. 
\frac{\partial \varphi (y,\lambda_q,\lambda_s,\gamma_s,
\ldots,\lambda_i,\ldots)}
{\partial \lambda_i}\right|_{\ldots=\lambda_i=\ldots=1}\;dy, 
\end{eqnarray} 
where $N_i^{thermal}$ stands for the number of particles of type $i$ 
coming directly from the collision. To this we must add the number of 
particles of the same species resulting through secondary production 
processes. Accordingly, the number $N_i^{theory}$, representing the 
theoretical prediction, is given by 
\begin{equation} 
N_i^{theory}= N_i^{thermal}+\sum_j b_{ij} N_j^{thermal}\;, 
\end{equation} 
where the $b_{ij}$ are are branching ratios corresponding to the decay of 
resonance $j$ into a particle of type $i$. 
 
As can be seen from (24), the system's volume enters in each particle 
multiplicity as a common multiplicative factor. This does not change 
if finite volume corrections are introduced to account for particle size, 
as in [14]. In such a case, the volume $V$ entering in (24) should be 
replaced by the free volume 
$\Delta$, given by 
\begin{equation} 
\Delta=\frac{V}{1+\varepsilon_{pt}(T,\{\lambda\})/4B}\;. 
\end{equation} 
In the above relation, $\varepsilon_{pt}$ is the energy density of a system 
of point particles which is a function of the {\it same} variables 
$(T,\{\lambda\})$ as those characterising the 
system with the extended particles ($\{\lambda\}$ denotes all fugacities 
collectively). 
 
The actual determination of $V$ itself is of no importance to our analysis. 
It will only be viewed as a parameter which serves to fit the data. So we 
need not consider volume corrections. Moreover, the analysis can 
be simplified (e.g. evaluation of the first and second derivatives with 
respect to the temperature) if we choose $VT^3/4\pi^3$ as our free 
parameter in the place of $V$. We shall, therefore, adopt this choice in 
the following. 
 
Our first task is to determine an optimal set of values for the 
thermal parameters, i.e. a collection ($VT^3/4\pi^3,T,\lambda_q, 
\lambda_s,\gamma_s$) which best fits the experimentally measured 
multiplicities. A primary concern is the imposition of the constraint 
$<S>=0$, which, according to the S-SBM scheme, amounts to enforcing 
eq. (7). This constraint does not allow, of course, the consideration of 
all the above parameters as being 
free (one of them has to be determined by the rest). Our way of ensuring 
the constraint is by introducing an extra parameter $l$ which plays the role 
of a Lagrange multiplier. 
 
Putting the above considerations together we form the 
function\footnote{We have denoted the function by $\chi^2$ for obvious 
reasons: It provides the measure of a chi-square fit.} 
\[\chi^2(VT^3/4\pi^3,T,\lambda_q,\lambda_s,\gamma_s,l)= 
\sum_{i=1}^N\left[\frac{N_i^{exp}-N_i^{theory} 
(VT^3/4\pi^3,T,\lambda_q,\lambda_s,\gamma_s)} 
{\sigma_i}\right]^2\] 
\begin{equation} 
+l\int_{\beta}^{\infty} x^3 \frac{\partial 
G(x,\lambda_q,\lambda_s,\gamma_s)} 
{\partial \lambda_s}dx\;\;, 
\end{equation} 
where $i$ runs over all hadrons measured in the 
experiment and $\sigma_i$ is the corresponding experimental error. Our 
aim will be to determine the values of the parameters\linebreak 
($VT^3/4\pi^3,T,\lambda_q,\lambda_s,\gamma_s,l$) which minimise the 
above function. The problem amounts to solving a system of six equations 
of the form 
\begin{equation} 
\frac{\partial \chi^2(x_1,\ldots,x_6)}{\partial x_i}=0 
\;\;(i=1,\ldots,6)\;, 
\end{equation} 
where the $x_i$ run over the set of parameters 
$(VT^3/4\pi^3,T,\lambda_q,\lambda_s,\gamma_s,l)$ 
 
The above system can be solved with a generalisation of the Newton-Raphson 
method to a multidimensional space. By this procedure we have to evaluate 
the second partial derivatives of $\chi^2$ with respect to its parameters. 
The method is quite sufficient when the point, which represents the optimised 
parameter values, is well inside the domain of the hadronic phase. On the 
other hand, it becomes very ineffective when this point is near the critical 
surface (or, worse, outside)\footnote{Recall that the SBM equations have 
analytic solutions only in the region inside the critical surface.}. 
This occurs because, when we begin to search for the solution with the 
Newton-Raphson method by giving an initial starting point, the subsequent 
points, through which the function passes during the evaluation procedure, 
in general oscillate. In the multidimensional space this 
oscillation may become very strong. When the optimal point happens to be 
near the critical surface it is quite likely that the aformentioned 
oscillations take us to points which lie on the outside. This will lead, 
of course, to a failure of the method. In view of this problem we have 
devised a different strategy of computation which {\it does} enable us to 
locate the desired minimum no matter how close it is to the critical surface. 
The relevant procedure is analyzed in detail in Appendix A. 
 
Another issue which is of importance to our analysis is whether the 
experimental data lead to an optimal point which lies inside or outside 
the critical surface. It would be of great use if we could know beforehand, so 
that we do not try to reach an elusive ``inside'' point. To this end, we have 
developed (see Appendix B) a method through which we can determine 
whether the minimum lies inside our outside the critical surface without 
actually evaluating its exact location. We can also estimate what percentage 
of the region, as defined by the experimental errors of the data (in particle 
ratios), is inside and what outside the critical surface. 
 
A final matter of methodological concern pertains to the relevance 
of Bose/Fermi statistics effects and the impact they might have on our 
analysis. As it turns out (see next section) the source of greatest worry, 
as far as discrepancies between predicted and 
experimental values are concerned, are the (negative) pion multiplicities, 
in particular for the $S+S$ experiment. The problem is that the direct 
inclusion of quantum particle statistics is quite difficult to be 
accomodated by the SBM scheme [16], whose formulation is based on the 
Boltzmann approximation. So, in 
order to gain an approximate evaluation of the error that is due to the 
omission of the Bose/Fermi statistics we turn to the Ideal Hadron Gas 
(IHG), always with the use of the grand canonical ensemble. 
 
The method that we shall adopt consists of finding the optimised set of 
parameters ($VT^3/4\pi^3,T,\lambda_q,\lambda_s,\gamma_s$) from the S-SBM 
theory for the $S+S$ data, subject, of course, to the constraint $<S>_{S- 
SBM}=0$. Then we determine the set ($VT^3/4\pi^3,T,\lambda_q, 
\lambda_s',\gamma_s$) which represents an optimal point 
in the Ideal Hadron Gas (IHG), formulated in the Boltzmann 
approximation and in the grand canonical ensemble (a constraint of the 
form $<S>_{IHG}=0$ is naturally imposed). For the latter collection of 
parameters we calculate the particle multiplicities $(N_{IHG})_i$ that we 
are interested in. At this point we switch to Bose/Fermi statistics as per 
in the IHG. For this case (BF) and for the set of parameters 
($VT^3/4\pi^3,T,\lambda_q,\lambda_s'',\gamma_s$) which verify the 
constraint $<S>_{BF}=0$, we calculate, once again, the desired particle 
multiplicities $(N_{BF})_i$. From the last two evaluations it follows that 
the relevant error in the estimation of the $i$th multiplicity, when we 
neglect the quantum statistics in the IHG, is 
$f_i=[(N_{IHG})_i-(N_{BF})_i]/(N_{IHG})_i$. Our working assumption is that 
about the same amount of error persists in the case of the SBM. So any 
correction factor for this situation will be taken to be $\simeq 1+f_i$. This 
correction factor will only be used for the $h^-$ multiplicity in the $S+S$ 
experiment and, as it will be seen, is small (of the order of $3\%$).

\vspace{2cm} 
{\large {\bf 4. Analysis of the Experimental Data}} 
 
In this Section we shall present the results of the application of the 
methodology we just described to experimental data. Our primary emphasis 
will be placed on particle multiplicities extracted from the NA35 $S+S$ 
experiment at CERN. We shall, for purposes of checking the consistency of 
our scheme, also confront data from the $p+\overline{p}$ experiment UA5 at CERN. 
 
We shall display our results on multiplicities in a series 
of Tables where a $\chi^2$ fit with respect to the optimal set of 
thermodynamical parameters will be given as well. A series of 
the graphical presentations of our results will be 
depicted on the ($T,\mu_q$) plane. We 
shall first set $\gamma_s$ to its value determined from the 
$\chi^2$ fit. We also drop the dependence on the volume $V$, 
otherwise we would have to make a specific choice before being able to 
display our results on a two dimensional plane. Making our plots as 
independent as possible on fitted parameters, we shall turn to 
particle ratios. The ratios that will be used will be chosen via a procedure 
that will be described in Appendix C. For a given particle ratio $x$, with 
experimental value $x^{exp}$, we shall plot, for fixed $\gamma_s$, the
projections on the ($T,\mu_q$) plane of the lines
$x(T,\mu_q,\mu_s,\gamma_s)=x^{exp}+\delta x^{exp}$ and 
$x(T,\mu_q,\mu_s,\gamma_s)=x^{exp}-\delta x^{exp}$. This will be done 
for each the ratio. On the same figure we shall trace the 
projection on the ($T,\mu_q$) plane of the intersection of the surface 
$<S>=0$ with the critical surface (calculated for the same value of 
$\gamma_s$), which sets the limit of the hadronic domain. Finally, the 
fitted values of $T$ and $\mu_q$, which result from the multiplicities, will 
also be indicated, along with the errors that result from the $\chi^2$ fit, 
represented by a solid circle on each graph. In this way we can conclude 
whether all the bands that are formed from the ratios and their errors have a 
common overlapping region. Such an occurrence would verify that the measured 
multiplicities are consistent with a system in thermal and (partial) chemical 
equilibrium described by the S-SBM and would help us evaluate 
the corresponding $T$ and $\mu_q$. More interestingly, we can determine 
how close to the limit of the hadronic domain the common region of the 
particle ratios' bands is and to what extent that region lies either in 
part, or as a whole, outside this boundary. Finally, we shall be in 
position to compare the results from the $\chi^2$ fit and the 
graphical analysis from the particle ratios, so that we can evaluate the 
degree of their complementarity. 
 
In the case of $p+\overline{p}$ experiment similar graphs will also be 
given, with the difference that they will be depicted on the 
($T,\gamma_s$) plane, while no choice for one of the relevant parameters 
will be needed in order to draw the bands which correspond to the particle 
ratios.

\vspace{1cm} 
{\large{\bf 4.1. $\bf S+S$ experiment (NA35)}} 
 
The study of $S+S$ experimental data, which pertain to an isospin symmetric 
nucleus-nucleus colliding system, will be conducted first in the 
full phase space and subsequently at midrapidity. 
 
\vspace{1cm} 
{\large{\bf 4.1.1. Full phase space data}} 
 
We deal, first, with the multiplicities that cover the full phase space. 
The $4\pi$ data from the $S+S$ experiment at SPS (200 GeV/nucleon) [23] 
are listed in Table 1, along with experimental errors. 
With $B-\overline{B}$ we denote the $baryon-antibaryon$ multiplicity. 
In the present attempt we confront the full set of data which includes $h^-$ 
(mostly $\pi^-$) multiplicities. A separate analysis which excludes the 
$h^-$'s will also be conducted for reasons that will become obvious in the 
following. 
 
\vspace{0.5cm} 
{\bf 4.1.1.a. Analysis with the $\bf h^-$} 
 
In this analysis, the free phenomenological parameter of the model $H_0$ 
is fixed so that $T_0$ attains, for $\gamma_s=1$, its ``maximum'' value 
$T_0=183$ MeV (see [19]). The theoretically calculated multiplicities are 
listed in the third column of Table 1. For the theoretical estimate of the 
$h^-$ multiplicity, we have used a correction factor $f=1.0255$ to account 
for the effect of quantum statistics. The evaluation of this factor has 
already been elucidated in Section 3. 
 
The values of the optimised thermal parameters 
($VT^3/4\pi^3,T,\lambda_q,\gamma_s$) 
which are calculated from our $\chi^2$ fit 
are listed in the second column of Table 3, along with their errors. 
Here we have to note that in the collision of two $S$ nuclei it is 
not evident how many nucleons really interact, even for 
central collisions, so the baryon number is not 
known {\it a priori}. This can be seen from the $B-\overline{B}$ 
multiplicity in Table 1, which is the net baryon number per interaction. 
It is measured to be $54\pm3$, whereas the total baryon number of the two 
incident $S$ nuclei is 64. So $\mu_q$ has to be left as a free parameter to 
be determined by the fit. The aforementined set comprises the true 
collection of free parameters which 
remain on account of the $<S>=0$ constraint. Accordingly, the degrees of 
freedom (dof) of the $\chi^2$ fit are $9-4=5$, given that we have 9 
experimental points. The least value of $\chi^2$ which is achieved by this 
fit is also listed in the same column. For completeness we have included 
the value of $\lambda_s$, which does not belong to the independent set of 
parameters, as well as those of the chemical potentials $\mu_q$ and 
$\mu_s$, which result from the calculated fugacities. 
 
The analysis with the particle ratios is depicted in Fig. 5, where we have set 
$\gamma_s=0.664$. The chosen particle ratios are 
listed in Table 2, along with their experimental values and 
errors; they will be employed throughout our treatment of the full 
phase space data for the $S+S$ experiment. In Fig. 7 we depict 
the value of $(N_{exp}-N_{theory})/\sigma_{exp}$, that is the number of 
standard deviations between experimental and theoretical values. 
The particular way of displaying our results was so chosen because there is 
great difference in the order of magnitude of the measured multiplicities. 
The points of our fit are denoted by empty squares. 
 
\vspace{0.5cm}
\begin{center}
{\bf $\bf S+S$ (NA35) 4{\boldmath $\pi$} phase space} 
\begin{tabular}{|c|ccc|} \hline 
\hspace{0.3cm}Particles\hspace{0.3cm} & 
Experimental Data & Calculated with $h^-$ & 
Calculated without $h^-$ \\ 
&&&(Case B)\\ \hline\hline 
$K^+$ & $12.5\pm 0.4$ & 12.581 & 12.706 \\ 
$K^-$ & $6.9\pm 0.4$ & 7.4590 & 6.6332 \\ 
${K_s}^0$ & $10.5\pm 1.7$ & 9.8106 & 9.3791 \\ 
$\Lambda$ & $9.4\pm 1.0$ & 7.8385 & 9.6767 \\ 
$\overline{\Lambda}$ & $2.2\pm 0.4$ & 1.3720 & 2.0156 \\ 
$\overline{p}$ & $1.15\pm 0.40$ & 1.9994 & 1.5117 \\ 
$p-\overline{p}$ & $21.2\pm 1.3$ & 22.849 & 21.529 \\ 
$B-\overline{B}$ & $54\pm 3$ & 53.544 & 52.348 \\ 
$h^-$ & $98\pm 3$ & 94.086$^{\rm a}$ & 71.227$^b$ \\ \hline 
\end{tabular} 
\end{center} 
 
{\footnotesize 
$^{\rm a}$ A correction factor 1.0255 has been included for the effect of 
Bose statistics. 
 
$^b$ A correction factor 1.0171 has been included for the effect of Bose 
statistics. This multiplicity is not included in the fit.} 
 
\begin{center} 
Table 1. Experimentally measured full phase space multiplicities in the 
NA35 $S+S$ experiment and their theoretically fitted values by S-SBM, 
with the inclusion of the $h^-$ multiplicity and without it (case B). 
\end{center} 
\vspace{0.5cm} 
 
It is evident that the fit, where all the multiplicities are included, is 
not so good, as can be witnessed from the relatively large value of 
$\chi^2$ (16.73, with 5 degrees of freedom). This is similar to the result of 
Becattini ($\chi^2/dof=17.2/5$) [9] and of Sollfrank 
($\chi^2/dof=11.6/4$) [10], who have conducted a fit to the multiplicity 
data from NA35 $S+S$ (with $h^-$ included) for the ideal hadron gas case, 
formulated in the canonical ensemble. 
 
Looking at Fig. 5 one can see that 
there is no common overlapping region between the particle ratios. The 
fitted value for the pair $(T,\mu_q)$ seems to 
lie inside the hadronic phase, {\it albeit} near its 
limits. An analysis conducted along the lines described in Appendix B 
leads, for the values of particle ratios listed in Table 2, 
to a probability of $92.97\%$ (238 points out of 256) for lying 
inside\footnote{This probability will be refered to as $P_{INSIDE}$ and its 
value will be listed to the appropriate Tables.} and 
$7.03\%$ (18 points out of 256) for lying outside the hadronic phase. 
Another point of note is that the strange 
sector is greatly suppressed, leading to $\gamma_s=0.66$. 

\vspace{0.5cm}
\begin{center}
{\bf $\bf S+S$ (NA35) 4{\boldmath $\pi$} phase space} 
\begin{tabular}{|c|cc|} \hline 
Particle Ratios used & Experimental Values & 
Calculated without $h^-$ \\ 
&&(Case A)\\ \hline\hline 
$K^-/K^+$ & $0.552\pm 0.037$ & 0.52811 \\ 
${K_s}^0/K^+$ & $0.84\pm 0.14$ & 0.74268 \\ 
$\Lambda/K^+$ & $0.752\pm 0.084$ & 0.73901 \\ 
$\overline{\Lambda}/K^+$ & $0.176\pm 0.032$ & 0.14413 \\ 
$\overline{p}/K^+$ & $0.092\pm 0.032$ & 0.12044 \\ 
$p-\overline{p}/K^+$ & $1.70\pm 0.12$ & 1.7191 \\ 
$B-\overline{B}/K^+$ & $4.32\pm 0.28$ & 4.1593 \\ 
$h^-/K^+$ & $7.84\pm 0.35^c$ & 5.8036$^d$ \\ \hline 
\end{tabular} 
\end{center} 
 
{\footnotesize 
$^c$ This ratio is not included in the fit. 
 
$^d$ A correction factor 1.0199 has been included for the effect of Bose 
statistics.} 
 
\begin{center} 
Table 2. Particle ratios from the experimentally measured full phase space 
multiplicities in the $S+S$ experiment and their theoretically fitted 
values by S-SBM, without the inclusion of the $h^-/K^+$ particle ratio 
(case A). 
\end{center} 
 
\vspace{0.5cm} 
{\bf 4.1.1.b. Analysis with the $\bf h^-$ excluded} 
 
We now perform the same analysis, but without explicit input from the 
$h^-$ multiplicity. The latter can, of course, be calculated 
theoretically but with the optimal parameter set which results from the 
fit to the rest of the multiplicities. It is now determined that, for the same 
choice of $H_0$, the values of the thermal parameters lead {\bf outside} the 
hadronic domain. Not being able to locate the point where $\chi^2$ 
attains its absolute minimum in this case, we search for that point on the 
critical surface which optimizes the value of $\chi^2$. Clearly, this is the 
closest we can get to the absolute minimum. In order to carry out such an 
analysis we have to turn 
to the method described in Appendix B. We designate this procedure as 
case A. The theoretically calculated particle ratios are listed in the 
last column of Table 2 and are plotted with empty triangles in Fig. 
8. The parameters that represent the location of the aforementioned least 
value of $\chi^2$ are listed in the second column of Table 3. Values for 
the volume have not been entered in this column, since particle ratios 
were used in the analysis. Note also that the fitted parameters are 
not accompanied by errors, which cannot be evaluated due to the 
fact that we are not at the absolute minimum value of $\chi^2$. In Fig. 6 
we have plotted the bands for the particle ratios that have been employed in 
the analysis, while the single point designates the location of the least 
value of $\chi^2$ in the hadronic phase (on the critical surface).

\vspace{0.5cm}
\begin{center}
{\bf $\bf S+S$ (NA35) $4\pi$ phase space} 
\begin{tabular}{|c|ccc|} \hline 
Fitted Parameters & Fitted with $h^-$ & 
Fitted without $h^-$ & Fitted without $h^-$ \\ 
&&(Case A)&(Case B)\\ \hline\hline 
$T$ (MeV) & $169.7\pm 9.1$ & 176.7 & $184.3\pm5.1$ \\ 
$\lambda_q$ & $1.538\pm 0.064$ & 1.609 & $1.613\pm0.079$ \\ 
$\gamma_s$ & $0.664\pm 0.056$ & 0.909 & $0.96\pm0.16$ \\ 
$VT^3/4\pi^3$ & $1.61\pm 0.51$ & $-$ & $0.61\pm0.31$ \\ 
$\chi^2/dof$ & $16.73\;/\;5$ & $3.06\;/\;4^{\;e}$ &$2.62\;/\;4$ \\ 
$\lambda_s$ & $1.089\pm 0.056$ & 1.010 & $0.989\pm0.052$ \\ 
$\mu_q$ (MeV) & $73.1\pm8.0$ & 84.0 & $88.2\pm9.3$ \\ 
$\mu_s$ (MeV) & $14.4\pm8.8$ & 1.7 & $-1.9\pm9.7$ \\ 
$P_{INSIDE}$ & $92.97\%$ & $25.78\%$ & $-$ \\ \hline 
\end{tabular} 
\end{center} 
 
{\footnotesize 
$^e$ It is the minimum of $\chi^2$ within the Hadron Gas with $T_0=183$ 
MeV (for $\gamma_s=1$), not the absolute minimum.} 
 
\begin{center} 
Table 3. Results of the analysis by S-SBM of the experimental 
data from $S+S$ experiment ($4\pi$ phase space), with the inclusion of the 
$h^-$ multiplicity and without it (cases A and B). 
\end {center}

\vspace{0.5cm} 
What {\it can} be seen from the last analysis is that the quality of the 
fit is very good. The value of $\chi^2$ (though not its absolute minimum) 
is very small (3.06 with 4 degrees of freedom), compared to the previous 
one. All the calculated particle ratios that enter the fit are in very good 
agreement with the experimental values, as can be seen from the second 
column of Table 2 and Fig. 8. The last ratio, however, which includes the 
$h^-$ multiplicity is very far from its experimental measurement. One can 
draw the conclusion that, for the thermal parameters coming out of this fit, 
the theoretically calculated number for the $h^-$'s (mostly $\pi^-$'s) is 
much less than the experimentally recorded one. 
 
From Fig. 6 one can see that bands for all the particle ratios, except for the 
one with the $h^-$ multiplicity, converge to the optimal point. If the 
hadronic phase were allowed 
to occupy more space, one could infer that a good overlapping region 
would be formed. We evaluated the probability of 
being outside and inside the critical surface. It turns out that there 
is a $25.78\%$ chance for being inside (33 points out of 128) and 
$74.22\%$ for being outside (95 points out of 128). Finally, one notices that 
the strange sector is almost fully saturated, as $\gamma_s=0.909$. 
 
To complete the analysis without the $h^-$'s we perform another kind of 
fit, referred to as case B. This time we allow for the hadronic phase to 
occupy more space, so that the point of absolute minimum $\chi^2$ falls 
inside this phase. Such an arrangement can be achieved via the unphysical 
situation wherein $H_0$ is set so that $T_0=193$ MeV\footnote{With this 
value the absolute minimum is just enclosed in the hadronic phase.} for 
$\gamma_s=1$. From the relevant study we can gain a better feeling of the 
quality of the fit without $h^-$ and to what extend full strangeness is 
attained. Our results indicate that when we vary $H_0$, which amounts to 
shifting the critical surface, the minimum value of $\chi^2$ remains 
basically constant. All the fitted thermal parameters, except for $T$ and 
$VT^3/4\pi^3$, undergo very slight changes. Our numerical results for the fit 
of case B give $\chi^2/dof=2.62/4$ and $\gamma_s=0.96\pm0.16$, i.e. the 
strange sector is fully saturated. 
The calculated values of the multiplicities are listed in Table 1 and plotted 
as empty triangles in Fig. 7. One notices again the difference between the 
experimental and the theoretical values of $h^-$. The optimised parameters 
for this fit are listed in Table 3 (last column).

\newpage
{\large{\bf 4.1.2. Midrapidity data}} 
 
We now turn to the analysis of multiplicity data restricted to the 
midrapidity region. The data are taken from Ref. [23,3] and are listed in 
Table 4. The rapidity interval is $2<y<3$ for all multiplicities, except for 
the $\overline{p}$ which are measured in the interval $3<y<4$. This poses no 
problem since the system is symmetric around $y=3$. 
 
\vspace{0.5cm}
\begin{center}
{\bf $\bf S+S$ (NA35) midrapidity} 
\begin{tabular}{|c|ccc|} \hline 
\hspace{0.3cm}Particles\hspace{0.3cm} & 
Experimental Data & Calculated with $h^-$ & 
Calculated without $h^-$ \\ \hline\hline 
$K^+$ & $3.2\pm 0.5$ & 3.7659 & 3.2874 \\ 
$K^-$ & $2.2\pm 0.5$ & 2.4957 & 2.0594 \\ 
$\Lambda$ & $2.05\pm 0.2$ & 1.9184 & 2.0593 \\ 
$\overline{\Lambda}$ & $0.57\pm0.2$ & 0.33815 & 0.53988 \\ 
$\overline{p}$ & $0.4\pm 0.1$ & 0.47255 & 0.41052 \\ 
$p-\overline{p}$ & $3.2\pm 1.0$ & 4.3518 & 3.0375 \\ 
$h^-$ & $26\pm 1$ & 25.544$^f$ & 16.617$^g$ \\ \hline 
\end{tabular} 
\end{center} 
 
{\footnotesize 
$^f$ A correction factor 1.031 has been included for the effect of 
Bose statistics. 
 
$^g$ A correction factor 1.021 has been included for the effect of Bose 
statistics. This multiplicity is not used in the fit.} 
 
\begin{center} 
Table 4. Experimentally measured midrapidity multiplicities in the $S+S$ 
experiment and their theoretically fitted values by S-SBM, with and 
without the inclusion of the $h^-$ multiplicity. 
\end{center}

\vspace{0.5cm}
{\bf 4.1.2.a. Analysis with the $\bf h^-$ included} 
 
Once again we perform a $\chi^2$ fit with the $h^-$ included. The theoretical 
multiplicities are listed in Table 4 and plotted in Fig. 11. For the $h^-$ a 
correction factor $f=1.031$ has been used to account for the effect of the 
Bose/Fermi statistics. The optimal set of values for the thermal parameters 
of this fit are listed in Table 6. The quality of the fit is, once again, not 
so good: $\chi^2/dof=5.59/3$. This also becomes evident from 
Fig. 9, where we have plotted the ratios. From 
this graph we can see that there is no common overlapping area of the 
particle ratios. The location of the minimum of $\chi^2$ is observed to be 
well inside the hadronic domain. We actually find $100\%$ probability for the 
point representing the experiment to be inside the hadronic sector 
(64 points are inside out of 64). Strangeness is also found suppressed, as 
$\gamma_s=0.78\pm0.12$. 
 
\vspace{0.5cm}
\begin{center}
{\bf $\bf S+S$ (NA35) midrapidity} 
 
\begin{tabular}{|c|c||c|c|} \hline 
Particle Ratios used&Experimental&Particle Ratios used&Experimental \\ 
for the fit with $h^-$&Values&for the fit without $h^-$ &Values \\ 
\hline\hline 
$K^+/h^-$& $0.123\pm 0.020$ &$K^+/\Lambda$& $1.56\pm0.29$ \\ 
$K^-/h^-$& $0.0846\pm 0.020$ &$K^-/\Lambda$& $1.07\pm0.27$ \\ 
$\Lambda/h^-$& $0.0788\pm 0.0083$ &$\overline{\Lambda}/\Lambda$ & 
$0.28\pm0.10$ \\ 
$\overline{\Lambda}/h^-$& $0.0219\pm 0.0077$ 
&$\overline{p}/\Lambda$ & $0.195\pm0.052$ \\ 
$\overline{p}/h^-$& $0.0154\pm 0.0039$ &$p-\overline{p}/\Lambda$& 
$1.56\pm0.51$ \\ 
$p-\overline{p}/h^-$& $0.123\pm 0.039$ &$h^-/\Lambda$& 
$12.7\pm1.3^g$ \\ \hline 
\end{tabular} 
\end{center} 
 
{\footnotesize $^g$ This ratio is not included in the fit.} 
 
\begin{center} 
Table 5. Particle ratios from the experimentally measured midrapidity 
multiplicities in the $S+S$ experiment, used in the analysis with and 
without the $h^-$ multiplicity. 
\end{center}

\vspace{0.5cm}
{\bf 4.1.2.b. Analysis with the $\bf h^-$ excluded} 
 
A different situation from the one described above arises when the $h^-$ 
are excluded. The results from the fit are listed in Tables 4 and 6, whilst 
the particle ratios that 
have been used are exhibited in Table 5. All the multiplicities, except $h^-$, 
are consistent with a system in thermal and chemical equilibrium as 
described by the S-SBM. This occurence can be directly inferred from 
the very small value of $\chi^2/dof=0.172/2$. 
 
In Fig. 10 we depict the situation for the particle ratios in the ($T,\mu_q$) 
plane. All the particle 
ratios, except the one that includes the $h^-$ multiplicity, form a common 
overlapping area which is partly within the hadronic domain. We find that 
the probability for a point belonging to this area to lie 
inside the hadron phase is $81.25\%$ (26 points are inside out of 32) and 
outside $18.75\%$ (6 points are outside out of 32). Note, also, that 
strangeness is almost oversaturated: $\gamma_s=1.15\pm0.26$. Of 
course, the error is such that the calculated $\gamma_s$ is consistent with 
the maximum possible value of $\gamma_s=1$. Note, 
on the other hand, that Sollfrank {\it et al.} in [3] have found 
$\gamma_s=1.19$ for the $S+S$ data at midrapidity. The study used the ideal 
hadron gas formulated in the grand canonical ensemble, with the cut to the 
input mass spectrum set at 2000 MeV. In our study, we have used as input all 
hadrons with mass up to 2400 MeV.

\vspace{0.5cm} 
\begin{center} 
{\bf $\bf S+S$ (NA35) midrapidity}\\ 
\begin{tabular}{|c|cc|} \hline 
Fitted Parameters 
&\hspace{0.5cm}Fitted with $h^-$\hspace{0.5cm} 
&\hspace{0.5cm}Fitted without $h^-$\hspace{0.5cm} \\ \hline\hline 
$T$ (MeV) & $156.7\pm 8.1$ &$171.7\pm9.4$ \\ 
$\lambda_q$ & $1.485\pm 0.081$& $1.450\pm0.075$ \\ 
$\gamma_s$ & $0.78\pm 0.12$ & $1.15\pm0.26$ \\ 
$VT^3/4\pi^3$ & $0.62\pm 0.15$ & $0.22\pm0.13$ \\ 
$\chi^2/dof$ & $5.59\;/\;3$ &$0.172\;/\;2$ \\ 
$\lambda_s$ & $1.149\pm 0.050$ & $1.045\pm0.055$ \\ 
$\mu_q$ (MeV) & $61.9\pm9.2$ & $63.8\pm9.6$ \\ 
$\mu_s$ (MeV) & $21.7\pm6.9$ & $7.6\pm9.0$ \\ 
$P_{INSIDE}$ & $100\%$ & $81.25\%$ \\ \hline 
\end{tabular} 
\end{center} 
 
\begin{center} 
Table 6. Results of the S-SBM analysis of the NA35 $S+S$ experiment 
(midrapidity), with and without the inclusion of the $h^-$ multiplicity. 
\end {center}

\vspace{1cm} 
{\large{\bf 4.1.3. General observations on the $\bf S+S$ data analysis}} 
 
Some general remarks are in order. Let us start by commenting on the 
correction factors due to Bose/Fermi 
statistics. This correction is worth taking into 
account only for the $h^-$ multiplicity. To demonstrate our claim, we 
consider the magnitude of the error of omitting the correct 
statistics by evaluating it for each multiplicity taken from the $S- 
S$ experiment. Moreover, we choose that set of parameters which leads to 
the greatest correction factor. The results appear in Table 7. 
 
\vspace{0.5cm}
\begin{center}
\begin{tabular}{|c|c|} \hline 
Particles ($i$)& 
$\frac{(N_{IHG})_i-(N_{BF})_i}{(N_{IHG})_i}$ ($\%$)\\ \hline\hline 
$K^+$ & $0.510$ \\ 
$K^-$ & $0.578$ \\ 
${K_s}^0$ & $0.527$ \\ 
$\Lambda$ & $0.228$ \\ 
$\overline{\Lambda}$ & $-0.273$ \\ 
$\overline{p}$ & $-0.034$ \\ 
$p-\overline{p}$ & $-0.092$ \\ 
$B-\overline{B}$ & $-0.017$ \\ 
$h^-$ & $3.043$ \\ \hline 
\end{tabular} 
\end{center} 
 
\begin{center} 
Table 7. Evaluation of the largest error in the analysis of the $S+S$ data 
due to the omission of the Bose/Fermi statistics. In the evaluation reference 
to the Ideal Hadron Gas model is made. 
\end {center} 
 
\vspace{0.5cm} 
The analysis for both the full phase space and midrapidity give a
fit which is not so good when the $h^-$ are included, 
whilst all other multiplicities are very well fitted when the $h^-$ 
are excluded. We conclude that the thermal parameters (temperature, 
chemical potentials and $\gamma_s$) are more accurate when 
evaluated without the $h^-$ multiplicity. The measured $\pi^-$ 
from the experiment are found to be, in both cases, a lot more compared to 
the theoretical predictions. The strangeness suppression factor 
$\gamma_s$ is also near its maximum value 1 in the absence of the $h^-$ 
multiplicity. 
 
Next, we comment on the fact that the fitted temperature consistently drops
when the $h^-$ are included in the analysis. One may think that this 
should not happen, since the increase in temperature should lead to the 
increase of particle multiplicity. This is true, of course, provided the 
volume of the system is held fixed. In the fit we have performed the volume 
is not known; it simply enters as a free parameter. 
Thus what really is important is not the actual but the relative magnitude of 
the different multiplicities. When the $h^-$ 
are increased much more than the other multiplicities, then the ratios of the 
latter over $h^-$ are decreased. Such a situation 
corresponds to lower temperature and probably greater volume (which, we 
repeat, was irrelevant to our analysis). 
 
A final remark concerns the fact that at midrapidity the evaluated 
quarkchemical potential $\mu_q$ is found to be somewhat less compared to 
full phase space case. This is expected, because the midrapidity region has 
smaller baryon number than the fragmentation region [23] (The full baryon 
number in the latter case {\it is} included in the $4\pi$ region). What is 
not so expected is that the fitted temperatures are lower at midrapidity than 
in $4\pi$ rapidity regime. This can be explained by the difference that 
exists in the fitted values of $\gamma_s$. The temperature and $\gamma_s$ are 
strongly correlated: increase in $\gamma_s$ leads to decrease of temperature 
and vice versa (see Fig. 2). The fitted $\gamma_s$ is found to be greater at 
midrapidity and that leads to a drop of the fitted temperature. Had we, on 
the other hand, calculated the temperature in full phase space 
and at midrapidity with $\gamma_s$ fixed to the same value for both cases, 
then we would have found greater temperature for the midrapidity case. 
 
\vspace{1cm} 
{\large{\bf 4.2. $\bf p+\overline{p}$ experiment (UA5)}} 
 
We shall, now, concern ourselves with another isospin symmetric system, 
namely $p+\overline{p}$ collisions. An analysis of the multiplicity data of 
the relevant UA5 experiment has been performed by Becattini in [8], based on 
a canonical formulation of the ideal hadron gas. Our analysis will be 
performed through the grand canonical formulation of the S-SBM. The data we 
shall use are taken from [24,8] and are listed in Table 8, for the different 
center of mass energies. 
 
The examination of this system calls for a qualitatively different approach 
as compared to the $S+S$ case. In the present situation, a $p$ always 
interacts with a $\overline{p}$. Thus the total baryon number is identically 
zero. We can thereby set the additional constraint on our system specified by 
$<B>=0$. The $<S>=0$ constraint still exists, as well. These two equations 
can easily be solved analytically for the fugacities and their solution gives 
\begin{equation} 
\lambda_q=\lambda_s=1\Leftrightarrow\mu_q=\mu_s=0\;\;, 
\end{equation} 
no matter what the value of $\gamma_s$ is. So we are only left with 3 free 
parameters to evaluate from the fit and no constraint to fulfill. These 
parameters are ($VT^3/4\pi^3,T,\gamma_s$). In a similar manner as in 
the previous subsections, we perform a $\chi^2$ fit. In Table 8 we list the 
theoretically calculated multiplicities for the different energies and have 
plotted them in Fig. 15. In Table 8 we also list the ratios that we depict on 
the ($T,\gamma_s$) plane in Figs. 12, 13 and 14. Our fitted parameters are 
tabulated in Table 9. 

\vspace{0.5cm} 
\begin{center} 
{\bf $\bf p+\overline{p}$ (UA5) 4{\boldmath $\pi$} phase space} 
\begin{tabular}{|c|cc||c|c|} \hline 
Particles & Experimental& Calculated&Particle&Experimental \\ 
 &Data& &Ratios used&Values \\ \hline\hline 
\multicolumn{5}{l}{$\sqrt{s}=200$ GeV} \\ \hline 
$Charged$ & $21.4\pm 0.4$ & 21.379 & 
${K_s}^0/{\scriptstyle Charged}$ & $0.0350\pm 0.0043$ \\ 
${K_s}^0$ & $0.75\pm 0.09$ & 0.77497 & 
$n/{\scriptstyle Charged}$ & $0.0350\pm 0.0047$ \\ 
$n$ & $0.75\pm 0.10$ & 0.77828 & 
$\Lambda/{\scriptstyle Charged}$ & $0.0107\pm 0.0028$ \\ 
$\Lambda$ & $0.23\pm0.06$ & 0.19934 & 
$\Xi^-/{\scriptstyle Charged}$ & $0.00070\pm 0.00070$ \\ 
$\Xi^-$ & $0.015\pm 0.015$ &0.01251& & \\ \hline 
\multicolumn{5}{l}{$\sqrt{s}=546$ GeV}\\ \hline 
$Charged$ & $29.4\pm 0.3$ & 29.392 & 
${K_s}^0/{\scriptstyle Charged}$ & $0.0381\pm 0.0027$ \\ 
${K_s}^0$ & $1.12\pm 0.08$ & 1.1325 & 
$\Lambda/{\scriptstyle Charged}$ & $0.0090\pm 0.0019$ \\ 
$\Lambda$ & $0.265\pm0.055$ & 0.29660 & 
$\Xi^-/{\scriptstyle Charged}$ & $0.00170\pm 0.00051$ \\ 
$\Xi^-$ & $0.050\pm 0.015$ &0.01984& & \\ \hline 
\multicolumn{5}{l}{$\sqrt{s}=900$ GeV}\\ \hline 
$Charged$ & $35.6\pm 0.9$ & 35.472 & 
${K_s}^0/{\scriptstyle Charged}$ & $0.0385\pm 0.0038$ \\ 
${K_s}^0$ & $1.37\pm 0.13$ & 1.4288 & 
$n/{\scriptstyle Charged}$ & $0.0281\pm 0.0057$ \\ 
$n$ & $1.0\pm 0.2$ & 1.1426 & 
$\Lambda/{\scriptstyle Charged}$ & $0.0107\pm 0.0023$ \\ 
$\Lambda$ & $0.38\pm0.08$ & 0.32629 & 
$\Xi^-/{\scriptstyle Charged}$ & $0.00098\pm 0.00056$ \\ 
$\Xi^-$ & $0.035\pm 0.020$ & 0.02236 & &\\ \hline 
\end{tabular} 
\end{center} 
 
\begin{center} 
Table 8. Experimentally measured in full phase space multiplicities 
and particle ratios from the $ p+\overline{p}$ experiment and 
corresponding theoretical predictions by S-SBM. 
\end{center} 

\vspace{0.5cm}
From Figs. 12-15 it can be inferred that a generally good fit exists. The 
minimum values of $\chi^2$ are similar to those found by Becattini in [8]. 
The overlapping region of the particle ratios is well inside the hadronic 
domain, as expected. The only discrepancy comes from the $\Xi^-$ multiplicity 
for $\sqrt{s}=546$ GeV which has an experimental value much greater than its 
theoretical prediction. This does not necessarily show something interesting. 
Finally, from the fitted parameters we can see that the temperature is 
almost constant with respect to the center of mass energy and that 
$\gamma_s$ appears to be almost constant as well. 

\vspace{0.5cm}
\begin{center} 
{\bf $\bf p+\overline{p}$ (UA5) 4{\boldmath $\pi$} phase space} \\ 
\begin{tabular}{|c|ccc|} \hline 
Fitted Parameters &$\sqrt{s}=200$ GeV 
&$\sqrt{s}=546$ GeV &$\sqrt{s}=900$ GeV \\ \hline\hline 
$T$ (MeV) & $159.0\pm 6.5$ &$159.3\pm6.3$ &$153.2\pm6.9$ \\ 
$\gamma_s$ & $0.422\pm 0.058$&$0.454\pm0.027$&$0.481\pm0.054$ \\ 
$VT^3/4\pi^3$ & $0.296\pm 0.051$ & $0.398\pm0.064$ 
& $0.56\pm0.10$ \\ 
$\chi^2/dof$ & $0.448\;/\;2$ &$4.398\;/\;1$& $1.583\;/\;2$ \\ \hline 
\end{tabular} 
\end{center} 
 
\begin{center} 
Table 9. Results of the S-SBM analysis of the data from $p+\overline{p}$ 
experiment. 
\end {center}

\vspace{2cm} 
{\large{\bf 5. Assesments and Conclusions}} 
 
In this paper we have performed an S-SBM study of particle multiplicities, 
recorded in very high energy collisions in which isospin symmetry holds. On 
the theoretical side, we adopted the thermal description of the multiparticle 
system under {\it partial} chemical equilibrium conditions. 
 
The relevant costruction involved five thermodynamical parameters subject 
to the constraint $<S>=0$. Our first concern was to 
determine an optimal set of values for these parameters through a $\chi^2$ 
fit to the experimental particle multiplicities. This formed 
the basis of our numerical computations which had dual aim. On one hand 
to assess the viability of our thermal description and, on the 
other, to locate the source of the multi-hadron system, on the phase 
diagramme, i.e. inside or outside the hadronic domain. 
 
The most striking observation, pervaiding our overall analysis, 
concerns the role of (negative) pion multiplicities. A drastic change 
transpires, with respect to a $\chi^2$ fit of our results, between the 
cases with and without the inclusion of pions, irrespective of adjustments 
made for quantum particle statistics. Full chemical equilibration is 
consistently achieved when pions are {\it not} included. Typical is the case 
of the $4\pi$ $S+S$ data analysis which shows compatibility with a value for 
$\gamma_s$ close to unity, i.e. with complete strangeness saturation, when 
the pions are not considered, whereas $\gamma_s$ drops to a considerably 
lower value when they are. The implications from analysis involving particle 
ratios are similar. Consistency with an equilibrium 
thermal behaviour is far better when experimental input from the $h^-$ is 
excluded. A similar situation arises in thermal analyses based on ideal 
hadron gas models, see e.g. [10]. 
 
One {\it would} like to understand the situation better, e.g. whether it is 
the freeze-out process responsible, or it is a signal that the location of 
the source of the hadrons is outside the hadronic domain. 
The latter case brings out the one special feature of the S-SBM, namely that 
it presents a critical surface beyond which the 
hadronic phase gives its place to a new one. Our studies involving 
particle ratios have established criteria for identifying the possible 
location of sources for the multiparticle hadronic state {\it beyond} this
critical surface. The viability of such a possibility, assuming reasonable
values for the bag constant ($B^{1/4}=235$ MeV) and/or the
maximum critical temperature $T_0=183$ MeV, has come up in relation to 
our analysis of data from the $S+S$ experiment NA35 at CERN [23]. 
In particular, the case of the $4\pi$ particle ratios (excluding the $h^-$)
indicates that about 3/4 of the particle-emitting state resides beyond the
critical surface. This information, despite its semi-quantitative relevance,
cannot be overlooked or ignored. We evaluate it as a strong sign of reaching
and entering into the sought-for deconfined region beyond the hadron gas.

Perhaps a more quantitative way by which to address the same issue is through
entropy considerations. It was clear from our analysis that the $\pi^-$ 
measured by the experiment are much more than we calculated 
theoretically. Sollfrank in [10] has pointed out that these particles 
may originate from a high entropy phase [25-27]. 
 
In order to shed some light into the notable discrepancy between the 
observed high $h^-$ multiplicity in the $S+S$ data and the one predicted by 
S-SBM, we examine the entropy content, since it depends mostly on the $h^-$ 
multiplicity. We calculate this quantity for the thermodynamical 
parameters obtained from the fit to the $4\pi$ $S+S$ data, 
Table 3. We remind that this fit produces a minimum point just beyond the 
assumed limit of the hadronic phase ($T_0=183$ MeV, $B^{1/4}=235$ MeV). 
 
\vspace{0.5cm} 
\begin{center} 
\begin{tabular}{|c|c|} \hline 
&$\;\;\;\;$ Entropy $\;\;\;\;$ \\ \hline\hline 
S-SBM&341.73\\ \hline 
IHG&264.68\\ \hline 
QGP ($m_s=150$ MeV)&482.41\\ 
QGP ($m_s=300$ MeV)&464.36\\ 
QGP ($m_s=500$ MeV)&436.15\\ \hline 
\end{tabular} 
\end{center} 
 
\begin{center} 
Table 10. Calculated entropy for the fitted parameters of the $4\pi$ $S+S$ 
data, case B. 
\end {center} 

\vspace{0.5cm}
Table 10 presents the results of the respective calculations for the ideal
Hadron Gas, S-SBM and QGP
formalisms. We note that the S-SBM value is within $(71-78)\%$ of the QGP 
one for strange quark mass in the range $150-300$ MeV and 
tends towards it. In contrast, the IHG value is within $(55-57)\%$ of the QGP 
value and about $77\%$ of the S-SBM one. Noting that the predicted to the 
observed $h^-$ multiplicity is about $73\%$ of the experimentally recorded 
value, we may infer that the $S+S$ interaction at 200 GeV/nucleon may have
just (initiated) a deconfinement phase transition and the $h^-$ excess comes
from an early stage of a deconfined quark matter state\footnote{We have
suggested in [5] that, between the hadron phase and the ideal QGP one there is a {\bf
deconfined quark matter} phase with massive interacting quarks.}.
 
In future work we intend to extend the S-SBM even more, by including 
the electric charge fugacity, $\lambda_Q$. This will enable us to take fully 
into account the difference in the number of the $u$ and $d$ quark of the 
initial colliding nuclei. In this way the analysis of a greater 
variety of heavy-ion experiments, including $Pb+Pb$ collisions, will 
become possible.

\vspace{2cm} 
\begin{center} 
{\large{\bf Acknowledgement}} 
\end{center} 
 
{\it This work was partially supported by the programme $\rm \Pi ENE 
\Delta$ No. 1361.1674/31-1-95, Gen. Secretariat for Research and 
Technology, Hellas.}

\newpage 
{\large{\bf Appendix A}} 
 
As mentioned in Section 3, in performing our analysis we have to find a set of 
parameters ($VT^3/4\pi^3,T,\lambda_q,\lambda_s,\gamma_s$), in the 
context of the S-SBM, that corresponds to the minimum value of $\chi^2$. 
In doing so, through a multidimensional generalisation of the Newton-Raphson 
method, we come across oscillations of points produced by the method 
during the evaluation procedure, irrespective of whether these lie 
inside the critical surface. We are, therefore, in need of another method for 
locating the minimum of $\chi^2$ that fully takes into account the 
restriction of the region where our functions are analytically determined. Of 
course we have to locate the minimum under the condition that the constraint 
$<S>=0$ is fulfilled (recall that our way of fulfilling this constraint was 
by the introduction of an additional parameter which played the role of a 
Lagrange Multiplier). 
 
Due to the fact that the critical surface is defined by the equation 
$\varphi(T,\lambda_q,\lambda_s,\gamma_s)=\ln 4-1$, there is no 
particular value of one of the parameters through which we can tell when 
the critical surface is approached. A combination of values of all 
the parameters is needed. The solution to our problem calls for dropping 
one of the parameters in terms of which we have mapped the space enclosed by 
the critical surface and replacing it by $\varphi$. We choose to replace the 
temperature $T$. As far as the fulfilment of the $<S>=0$ 
constraint is concerned, we propose to try to find the minimum without 
ever leaving the $<S>=0$ surface. That means that one more parameter will no 
longer be considered as free, preferably $\lambda_s$. Its value at every 
point will be determined by the rest of the parameters. 
 
The method can be summarised as follows. From the original set of 
parameters we choose to describe $\chi^2$ as a function of 
($VT^3/4\pi^3,\varphi,\lambda_q,\gamma_s$) when multiplicities are 
the experimental input and of ($\varphi,\lambda_q,\gamma_s$) when 
the input are the ratios. In the following we consider the case of the 
ratios. We start by an initial point which is inside the critical surface, 
denoted by ($\varphi^0,\lambda_q^0,\gamma_s^0$). Such a point is easy to 
specify: it is enough for it to satisfy $\varphi^0\le \ln 4-1$. 
In order to determine the value of $\chi^2$ at this initial point we have to 
find the corresponding fugacity $\lambda_s$. For this reason we solve the 
system of the two equations: 
\begin{equation} 
\varphi(T,\lambda_q^0,\lambda_s,\gamma_s^0)=\varphi^0 
\end{equation} 
\begin{equation} 
<S>(T,\lambda_q^0,\lambda_s,\gamma_s^0)=0\;. 
\end{equation} 
The above system is solved numerically through a proper Newton-Raphson 
method. Let us stress that our designation of a starting point 
($T^{start},\lambda_s^{start}$) to the method is of paramount importance. If, 
again, we are led to points outside the critical surface then the present 
method will not solve our problem. To this end, it is very useful to observe 
that the critical surface around the value of $\mu_s=0$ is almost 
perpendicular to the ($T,\mu_q$) plane (see Fig. 9 in [19]). That means that 
whether we are outside or inside the critical surface will be determined by 
the rest of parameters and not by $\lambda_s$. So it is not important what 
starting point we shall give to $\lambda_s$, e.g. a value 
$\lambda_s^{start}=1$ will not cause 
any trouble. The starting value for the temperature, on the other hand, 
needs more caution. This can be given by solving numerically the equation: 
\begin{equation} 
\varphi(T^{start},\lambda_q^0,\lambda_s^{start},\gamma_s^0)=\varphi^0 
\;. 
\end{equation} 
Once the desired $T$ and $\lambda_s$ are specified $\chi^2$ can be 
subsequently evaluated. It is noted that the described routine has to be 
repeated every time the function $\chi^2$ is evaluated. 
 
The subsequent steps involve the performance of a number of one 
parameter minimisations. A lot of 
routines exist for this task (see for example [28]). We consider the function 
$\chi^2(\varphi,\lambda_q^1,\gamma_s^1)$, which means that we hold 
fixed the parameters $\lambda_q=\lambda_q^1$ and $\gamma_s=\gamma_s^1$, 
and proceed with its minimisation. Suppose we find that its minimum is 
located at $\varphi=\varphi^2$. Then we minimise 
$\chi^2(\varphi^2,\lambda_q,\gamma_s^1)$ with respect to $\lambda_q$. 
Let $\lambda_q^2$ be the location of this minimum. We repeat the routine 
for $\chi^2(\varphi^2,\lambda_q^2,\gamma_s)$, this time with respect to 
$\gamma_s$, which gives us $\gamma_s^2$. Of course, the point 
($\varphi^2,\lambda_q^2,\gamma_s^2$) will not give the location of the 
absolute minimum, so that the whole procedure will have to be repeated until 
$\chi^2$ is efficiently reduced. Note that, during our trials to find minima 
for one of the parameters $\lambda_q$ or $\gamma_s$, we are not in danger of 
getting out of the critical surface because $\varphi$ is held fixed to a 
value less than $\ln4-1$. When we try to find a better $\varphi$, with 
$\lambda_q$ and $\gamma_s$ fixed, we can get out of the critical surface only 
if the method 
of one parameter minimisation gives a value to $\varphi$ greater than 
$\ln 4-1$. Usually, when this happens it means that the absolute minimum of 
$\chi^2$ is outside the critical surface. The method described in the 
following appendix will certify this fact. 
 
The method exhibited above is very slow compared to the generalised 
Newton-Raphson one, especially if the parameter $VT^3/4\pi^3$ is added (in the 
multiplicity case). But it {\it will} locate the desired minimum, even if it 
is near the critical surface, provided, of course, that it lies inside this 
surface. It can also be used to guide us near to the optimised parameters of 
the minimum, so the oscillation of the ``test'' points of Newton-Raphson is 
suppressed and we can reach the minimum with no danger of getting outside the 
region of analyticity.

\vspace{2cm} 
{\large{\bf Appendix B}} 
 
In this appendix we shall present the method which allows us to conclude 
whether a given set of experimental data leads to thermal parameters which 
define a point inside or outside the critical surface. 
Because this method is rather slow we shall use as experimental input 
particle ratios and not multiplicities, reducing, in this way, by one our free 
parameters. 
 
The method consists of, firstly, locating the minimum value of $\chi^2$ on 
the intersection of the critical surface with the $<S>=0$ surface. This 
intersection is no longer a single curve, as in [18,19]. Given that the new 
parameter $\gamma_s$ has been introduced into the S-SBM, the 
intersection is described by two variables which we choose to be 
($\lambda_q,\gamma_s$). For a given value of this pair the corresponding 
$\lambda_{s\;cr}$ can be found by numerically solving the generalisation 
of eq. (49) in [19]: 
\begin{equation} 
\int_1^0\frac{dz}{z-2}\cdot\left[ 
\frac{\frac{\textstyle \partial \varphi 
(y,\lambda_{q\;cr},\lambda_{s\;cr},\gamma_{s\;cr})} 
{\textstyle \partial \lambda_s}} 
{y^5 \cdot\frac{\textstyle 
\partial \varphi (y,\lambda_{q\;cr},\lambda_{s\;cr},\gamma_{s\;cr})} 
{\textstyle \partial y}} \right]_ 
{\textstyle z=2-\exp 
[G(y,\lambda_{q\;cr},\lambda_{s\;cr},\gamma_{s\;cr})]}\;\;=\;0\;\;. 
\end{equation} 
Then the value of multiplicities prior to decays, which enter the theoretical 
value of the ratios, is given by (see eq. (21)): 
\[N_i^{thermal}(V,T_{cr},\lambda_{q\;cr},\lambda_{s\;cr}, 
\gamma_{s\;cr})=\frac{VT_{cr}^3}{4\pi^3 H_0}\cdot 
\int_1^0\frac{dz}{z-2}\cdot\] 
\begin{equation} 
\left[\frac{ \left. 
\frac{\textstyle \partial \varphi 
(y,\lambda_{q\;cr},\lambda_{s\;cr},\gamma_{s\;cr},\ldots,\lambda_i,\ldots)} 
{\textstyle \partial \lambda_i}\right|_{\ldots=\lambda_i=\ldots=1}} 
{y^5 \cdot\frac{\textstyle 
\partial \varphi (y,\lambda_{q\;cr},\lambda_{s\;cr},\gamma_{s\;cr})} 
{\textstyle \partial y}} \right]_ 
{\textstyle z=2-\exp 
[G(y,\lambda_{q\;cr},\lambda_{s\;cr},\gamma_{s\;cr})]}=0. 
\end{equation} 
Note that $V$ and $T_{cr}$ need not to be known since they cancel in the 
ratios. 
 
Knowing how to evaluate $\chi^2$ on the intersection we perform, as in 
Appendix A, as many steps as required, each of which consists of two 
succesive one parameter minimisations. In the end, we obtain a point 
$A^m=(\lambda_q^m,\gamma_s^m)$ which corresponds to the minimum 
value of $\chi^2$ on the aforementioned intersection. 
 
To proceed further we think that, if the absolute minimum of $\chi^2$ 
lies inside the critical surface, then by moving away from $A^m$ towards 
the interior there should be some direction where $\chi^2$ 
decreases. On the contrary, if the absolute minimum lies on the outside, 
the minimum value of $\chi^2$ within the hadronic 
domain should lie on its limits. Any move from this location, in any 
direction inside the critical surface, should lead to the increase of 
$\chi^2$. With these considerations in 
mind, we perform the following check. We form a grid of 9 points: 
$A^{ij}=(\lambda_q^m+\delta_i,\gamma_s^m+\delta_j)$, where 
$\delta_i=- 
\delta_1,0,\delta_1$ and $\delta_j=-\delta_2,0,\delta_2$, with $\delta_1$ 
and $\delta_2$ small numbers. The points $A^{ij}$ are taken on a surface very 
close to the critical one, which is almost parallel to it. Such a surface 
can be determined by $\varphi=\ln 4-1-\delta_3$, with $\delta_3$ a small 
positive number. The points $A^{ij}$ are located within the hadronic 
phase, very close to $A^m$ but in different directions from it. We evaluate 
$\chi^2$ for all $A^{ij}$. If {\it one} of the points $A^{ij}$ is 
found to lead to value of $\chi^2$ less than the one it has at 
$A^m$, then the absolute minimum of $\chi^2$ is considered to be inside the 
critical surface. If, on the contrary, {\it all} the points $A^{ij}$ possess 
greater values for $\chi^2$ than the one at $A^m$, then we surmise 
that the absolute minimum lies outside. 
 
With the method we just described we can evaluate the probability 
that an experimental point, with its experimental errors, is 
inside or outside the critical surface. Explicitly, 
suppose that we have $n$ ratios given by the experiment. 
One of these ratios, e.g. $x_1$, takes the experimental value 
$x_1^{exp}\pm\delta x_1^{exp}$ and so is somewhere in the interval $( 
x_1^{exp}-\delta x_1^{exp}, x_1^{exp}+\delta x_1^{exp})$. By the same 
token, a second ratio $x_2$ can lie in the interval $( x_2^{exp}-\delta 
x_2^{exp}, x_2^{exp}+\delta x_2^{exp})$ and so on. By taking one value 
for every ratio from its corresponding allowed interval, we form an 
$n$-valued point which is within the error margins of our data. This point 
can be fed as experimental input to $\chi^2$ and with the previous method we 
can find out whether it leads to an absolute minimum inside the hadron gas 
phase or not. We can, similarly, consider other points, forming different 
combinations among the allowed values of each ratio. The ratio of the number 
of points that are outside (inside)to the whole number of points considered 
give the probability that the data lead us outside (inside) the critical 
surface. 
 
In practice it takes a lot of time to process one point. So we only enter 
two possible values of every ratio at the limits of its interval. Thus the 
points that we consider as experimental input in $\chi^2$ are: 
\[(x_1^{exp}+i_1\delta x_1^{exp},x_2^{exp}+i_2\delta x_1^{exp},\ldots 
x_n^{exp}+i_n\delta x_n^{exp})\;,\] 
with $i_1=\pm1,i_2=\pm1,\ldots,i_n=\pm1$. This accounts to processing 
$2^n$ points for every experiment that provides $n$ particle ratios.

\vspace{2cm} 
{\large{\bf Appendix C}} 
 
In principle, it is equivalent to use as experimental input in a $\chi^2$ fit 
a number $N$ of measured particle multiplicities or a number of $N-1$ 
independent particle ratios. There are numerous combinations 
of $N-1$ ratios that can be formed from the same set of multiplicities. 
The way we proceed in practice is the following. 
After performing the $\chi^2$ fit with the multiplicities from 
the $S+S$ data, we conduct a number of $\chi^2$ fits using different sets 
of ratios. We found a great deal of difference among these results as both the 
optimised parameters and the minimum value of $\chi^2$ 
exhibited puzzling variations. 
 
Before we analyse the situation let us recall the basic reasons for which it 
{\it is} needed to deal with particle ratios. 
Firstly, to locate the possible overlapping region of the bands 
of the particle ratios drawn on the ($T,\mu_q$) plane. 
Secondly, to evaluate the least value of $\chi^2$ on the 
two-dimensional intersection of the critical surface with the $<S>=0$ 
surface when the absolute minimum lies outside the critical surface. 
Thirdly, to estimate the probability that the experimental 
measurements lead to an ``inside'' or an ``outside'' point. It follows that 
we cannot avoid facing the question which set of particle ratios to use 
in a way that it leads us to results as close as possible to those 
represented by the multiplicities. 
 
Translating the multiplicities to a set of ratios 
does not change anything as far as the centroid values are concerned. What 
{\it is} changed are the accompaning errors. The errors associated with 
the ratios depend on the way the multiplicities are coupled together to form 
them. To clarify this point, let us suppose that we choose the multiplicity 
$x_m$ with the greatest relevant error ($e_m=\delta x_m/x_m=max$) and couple 
it with all the rest of the multiplicities. Let us also suppose that among 
them exists a multiplicity 
with very small relevant error ($e_n=\delta x_n/x_n$). This multiplicity 
will now enter in the ratio $x_n/x_m$ and the relevant error will be 
$e_{nm}=\sqrt{e_n^2+e_m^2}\gg e_n$. When the multiplicity fit 
is performed, $x_n$ plays an important role in the 
evaluation of the fitted parameters, since it has a very narrow interval for 
allowed values. The $\chi^2$ fit has to arrange the free parameters so that 
$x_n^{theory}$ comes close to $x_n$ and the contribution to $\chi^2$ 
from this multiplicity is not too great\footnote{The dependence of 
$\chi^2$ on the relevant error can be seen if we arrange its terms as 
\newline $\chi^2=\sum_i 
\left(\frac{\textstyle 1-x_i^{theory}/x_i^{exp}} 
{\textstyle \sigma_i^{exp}/x_i^{exp}}\right)^2 
=\sum_i\left(\frac{\textstyle 1-x_i^{theory}/x_i^{exp}} 
{\textstyle e_i}\right)^2$.}. 
On the contrary, the big relevant error to the ratio that includes $x_n$ has 
the effect that the fit does not really take the smallness of $e_n$ too much 
into account. Therefore, in order for the results of the fit with the ratios 
not to come out too distorted, we have to pay attention not to 
upset the connection of every multiplicity to its error. One simple way to do 
this is to pick that multiplicity with the smallest relevant error and 
form ratios by coupling the rest of multiplicities with it. 
 
We have checked in practice that the fit to the ratios chosen with the 
abovementioned logic really lead to the same results (fitted parameters 
and value of $\chi^2$) as for the multiplicity fit. In the case of the $4\pi$ 
data from $S+S$ we find that the $K^+$ and the $h^-$ multiplicities have the 
smalest relevant error ($\simeq 3\%$). In order to have the same set of 
multiplicities to both of our fits (with and without $h^-$) we choose the 
$K^+$ and form their ratios with the rest. In the case of the midrapidity 
data from $S+S$ the $h^-$ has the smallest relevant error ($\simeq 3.8\%$), 
so it was chosen to form the ratios with the rest, in the case of the fit with 
the inclusion of $h^-$. When $h^-$ 
is excluded from the fit, the smallest error lies with $\Lambda$ ($\simeq 
9.8\%$) and thus this multiplicity was chosen. In the same way the 
$charged$ multiplicity was chosen for all the fits to $p+\overline{p}$ data.

\vspace{2cm} 
{\large{\bf Figure Captions}} 
\newtheorem{f}{Figure} 
\begin{f} 
\rm (a) Intersections of planes of constant s-quark chemical potential 
$\mu_s$ with the critical surface $\varphi(T,\mu_q,\mu_s,\gamma_s)= 
\ln 4-1$ for $T_0=183$ MeV (for $\gamma_s=1$).\\ 
(b) Intersections of planes of constant q-quark chemical potential 
$\mu_q$ with the critical surface $\varphi(T,\mu_q,\mu_s,\gamma_s)= 
\ln 4-1$ for $T_0=180$ MeV (for $\gamma_s=1$). 
\end{f} 
\begin{f} 
\rm Variation of the critical temperature, $T_0$, for zero chemical 
potentials $\mu_q$ and $\mu_s$ with $\gamma_s$. $T_0$ is set to be 183 
MeV when $\gamma_s=1$. 
\end{f} 
\begin{f} 
\rm Projection on the plane $(T,\mu_s)$ of intersections of planes of 
constant q-quark fugacity $\lambda_q$ ($\mu_q/T=0.4$) with the surface 
$<S>=0$ for two values of $\gamma_s$. $T_0$ is set to be 183 MeV when 
$\gamma_s=1$. 
\end{f} 
\begin{f} 
\rm (a) Variation of the projection on the plane $(\mu_q,\mu_s)$ of the 
intersection of the critical surface and the surface $<S>=0$ with 
$\gamma_s$, for two different values of $T_0$ (for $\gamma_s=1$). 
(b) Variation of the projection on the plane $(T,\mu_s)$ of the intersection 
of the critical surface and the surface $<S>=0$ with $\gamma_s$, for two 
different values of $T_0$ (for $\gamma_s=1$). 
\end{f} 
\begin{f} 
\rm Experimental Particle Ratios in the ($T,\mu_q$) plane for $S+S$ 
experiment measured in $4\pi$ phase space. $\gamma_s$ is set to 0.664. 
The point and the cross are from the $\chi^2$ fit with the $h^-$. The thick 
solid line represents the limits of the hadronic phase (HG) as they are set by 
S-SBM. The other line corresponding to the ratio ${K_s}^0/K^+$ lies at 
the domain of negative $\mu_q$, for $T>100$ MeV. 
\end{f} 
\begin{f} 
\rm Experimental Particle Ratios in the ($T,\mu_q$) plane for $S+S$ 
experiment measured in $4\pi$ phase space. $\gamma_s$ is set to 0.909. 
The point represented by the solid circle corresponds to the location of the 
least value within the hadron gas of $\chi^2$, 
without the $h^-$. The thick solid line represents the limits of the hadronic 
phase (HG) as they are set by S-SBM. The other line, pertaining to the 
ratio ${K_s}^0/K^+$, lies in the domain of negative $\mu_q$, for 
$T>100$ 
MeV. 
\end{f} 
\begin{f} 
\rm Comparison between the experimentally measured multiplicities in 
$4\pi$ phase space and the theoretically calculated values in the fit with 
$h^-$ and without $h^-$ (case B) for the $S+S$ experiment. The 
difference is measured in units of the relevant experimental error. 
\end{f} 
\begin{f} 
\rm Comparison between the experimentally measured particle ratios in 
$4\pi$ phase space and the theoretically calculated values without $h^-$ 
(case A) for the $S+S$ experiment. The difference is measured in units of 
the relevant experimental error. 
\end{f} 
\begin{f} 
\rm Experimental Particle Ratios in the ($T,\mu_q$) plane for $S+S$ 
experiment measured in midrapidity. $\gamma_s$ is set to 0.78. The point 
and the cross are from the $\chi^2$ fit with the $h^-$. The thick solid line 
represents the limits of the hadronic phase (HG) as they are set by S-SBM. 
\end{f} 
\begin{f} 
\rm Experimental Particle Ratios in the ($T,\mu_q$) plane for $S+S$ 
experiment measured in midrapidity. $\gamma_s$ is set to 1.15. The point 
and the cross are from the $\chi^2$ fit without the $h^-$. The thick solid 
line represents the limits of the hadronic phase (HG) as they are set by 
S-SBM. 
\end{f} 
\begin{f} 
\rm Comparison between the experimentally measured multiplicities in 
midrapidity and the theoretically calculated values in the fit with $h^-$ and 
without $h^-$ for the $S+S$ experiment. The difference is measured in units 
of the relevant experimental error. 
\end{f} 
\begin{f} 
\rm Experimental Particle Ratios in the ($T,\gamma_s$) plane for 
$p+\overline{p}$ experiment for $\sqrt{s}=200$ GeV measured in $4\pi$ 
phase space. The point represented by the solid circle corresponds to 
the $\chi^2$ fit. The thick solid line represents the limits of the hadronic 
phase (HG), as they are set by S-SBM. 
\end{f} 
\begin{f} 
\rm Experimental Particle Ratios in the ($T,\gamma_s$) plane for 
$p+\overline{p}$ experiment for $\sqrt{s}=546$ GeV measured in $4\pi$ 
phase space. The point represented by the solid circle corresponds to the 
$\chi^2$ fit. The thick solid line represents the limits of the hadronic 
phase (HG), as they are set by S-SBM. 
\end{f} 
\begin{f} 
\rm Experimental Particle Ratios in the ($T,\gamma_s$) plane for 
$p+\overline{p}$ experiment for $\sqrt{s}=900$ GeV measured in $4\pi$ 
phase space. The point represented by the solid circle corresponds to the 
$\chi^2$ fit. The thick solid line represents the limits of the hadronic 
phase (HG), as they are set by S-SBM. 
\end{f} 
\begin{f} 
\rm Comparison between the experimentally measured (solid circles) 
particle ratios in $4\pi$ phase space and the theoretically calculated values 
(empty squares) for the $p+\overline{p}$ 
experiment and for $\sqrt{s}=200$ GeV, 546 GeV and 900 GeV. The 
difference is measured in units of the relevant experimental error. 
\end{f}

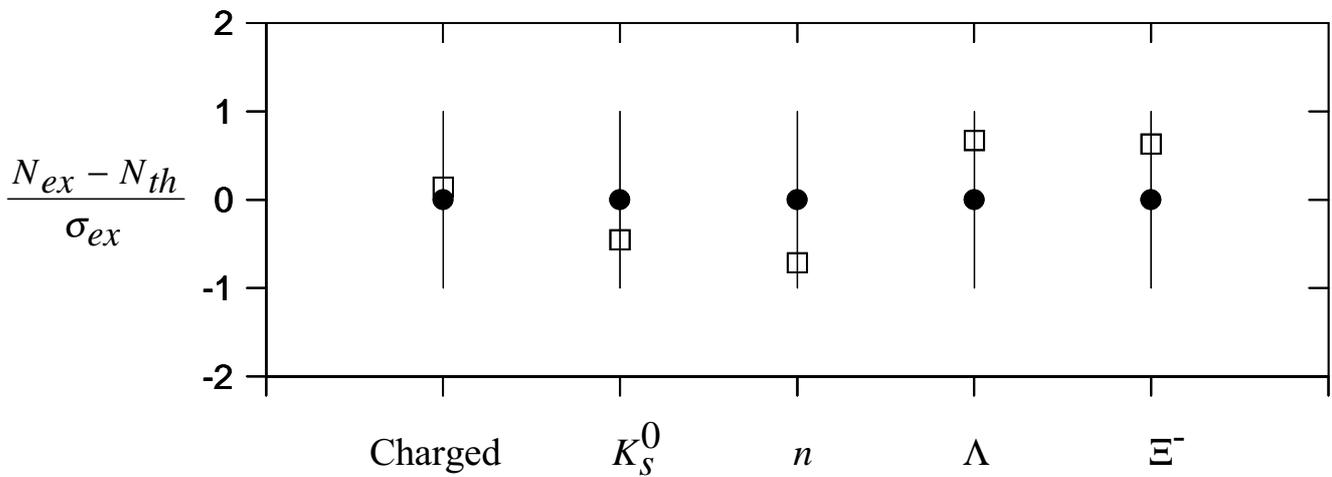
\end{document}